\documentclass[letter]{aa} 
\usepackage{graphicx}
\usepackage{txfonts}
\begin{document}
\title{
The host galaxy of the BL Lacertae object 
1ES 0647+250 and 
its imaging redshift}
\author{
J.K. Kotilainen\inst{1,2}
\and T. Hyv\"onen\inst{1}
\and R. Falomo\inst{3}
\and A. Treves\inst{4}
\and M. Uslenghi\inst{5}
}
\institute{
Finnish Centre for Astronomy with ESO (FINCA), University of Turku, V\"ais\"al\"antie 20, FI-21500 Piikki\"o, Finland
\and Tuorla Observatory, University of Turku, V\"ais\"al\"antie 20, FI-21500 Piikki\"o, Finland
\and Osservatorio Astronomico di Padova, INAF, vicolo dell'Osservatorio 5, 35122, Padova, Italy
\and Dipartimento di Fisica e Matematica, Universit\`a dell'Insubria, Via Valleggio 11, I-22100 Como, Italy, affiliated to INAF and INFN
\and INAF-IASF Milano, Via E. Bassini 15, Milano I-20133, Italy
}
\date{Received ... / Accepted ...}

  \abstract
{}
{
Since no spectroscopic redshift is available for the remarkable 
BL Lac object 1ES 0647+250, we aim to derive an estimate of its distance from 
the properties of its host galaxy.
}
{
We 
obtained
a deep, high-resolution near-infrared $H$-band image of 
the BL Lacertae object 1ES 0647+250.
} 
{
We are
able to detect the underlying host galaxy in 
the near--infrared. The host galaxy 
has an $H$-band magnitude 
of 16.9$\pm$0.2 and an effective radius of 1\farcs6$\pm$0\farcs3. 
Using the imaging redshift method by Sbarufatti et al. (2005), we estimate 
a redshift z = 0.41$\pm$0.06. 
This redshift is consistent with the previously reported imaging redshift 
estimate from the optical $i'$-band, z = 0.45$\pm$0.08 by 
Meisner \& Romani (2010), and with previously reported lower limits for 
the redshift. It is also in agreement with constraints from its 
gamma-ray emission. 
}
{
Imaging searches in the near-infrared, even with moderately sized telescopes, 
for the host galaxies of BL Lac objects at unknown redshift, are encouraged, 
as well as optical spectroscopy of 1ES 0647+250 with large telescopes to 
determine its spectroscopic redshift. 
}


\titlerunning{The host galaxy of 1ES 0647+250}

\maketitle

%

\section{Introduction}

In the study of extragalactic objects, a fundamental parameter is 
their distance. In active galactic nuclei (AGN), this is usually derived from 
the redshift z, measurable from emission lines, with the exception of 
BL Lac objects. Their practically featureless spectra are dominated by 
the Doppler-boosted emission of a relativistically beamed synchrotron jet, 
closely aligned with our line of sight (e.g. Urry \& Padovani, 1995). They are 
members of the high-energy family of AGN, showing strong 
nonthermal emission, superluminal motion and rapid/large variations in 
flux/polarization, and by definition, few (if any) very weak emission lines. 
The redshift of BL Lacs can, however, be derived from the absorption 
lines of their host galaxies. Despite this, even with deep spectroscopy at 
8-m class telescopes (e.g. Sbarufatti et al. 2006,2009), many BL Lacs remain 
without a spectroscopically determined redshift.

The ignorance about the redshift prevents us from constraining their 
high-energy physics, their bolometric power, multifrequency 
spectral energy distributions (SEDs) and synchrotron self-Compton (SSC) 
emission models. Most blazars (BL Lacs and flat spectrum radio quasars, FSRQ) 
emit the bulk of their power at MeV-GeV gamma-rays. Many blazars have been 
detected even at TeV gamma-rays using ground-based Cherenkov telescopes 
MAGIC, HESS, and VERITAS. Since VHE gamma-rays can be absorbed by 
the interaction with low energy photons of 
the extragalactic background light (EBL) via pair production, this opens 
the possibility of measuring the amount of EBL, which provides important 
information about the galaxy formation and star formation histories.  
The absorption depends strongly on the distance of the source and the energy 
of the gamma-rays. If the redshift of the source is known, the VHE gamma-ray 
spectrum can be used to derive limits on the EBL (e.g. Aharonian et al. 2006).  

The blazar 1ES 0647+250 was detected as a radio source in 
the MIT-Green Bank II survey at 5 GHz using the NRAO 91-m transit telescope 
(Langston et al. 1990). X-ray emission was discovered in the Einstein IPC 
Slew Survey (Elvis et al. 1992).
1ES 0647+250 was classified as a BL Lac object through X-ray/radio vs. 
X-ray/optical colour-colour diagrams and confirming optical spectroscopy by 
Schachter et al. (1993). Optical spectra taken at 1-2m class telescopes have 
validated
that 1ES 0647+250 has a featureless spectrum (Perlman et al. 1996). 
The object is highly variable with rapid variations 
observed from radio to X-ray bands (e.g. Giommi et al. 1995; 
Nieppola et al. 2006). The VLBA map of 1ES 0647+250 (Rector et al. 2003) 
does not show a distinct jet but there is some evidence for a faint, 
diffuse halo around the core, with weak evidence of elongation of the core to 
the north. VLA 6 cm observations (Perlman et al. 1996) of this source indicate 
a jet extending to the SW as well as a possible extension to the NW.  

The nucleus of 1ES 0647+250 is optically very bright, thus previous attempts 
to either determine its redshift spectroscopically or, until recently, 
to characterize its host galaxy have not been successful.
Thus, in spite of numerous variability studies of 
1ES 0647+250, it has never been possible to determine reliably e.g. the linear 
dimensions and luminosities of the varying components.

Upper limits to the TeV gamma-ray emission from this object have been reported 
by HEGRA (Aharonian et al. 2004) and by MAGIC (Abdo et al. 2009). 
Recently, there was a tentative detection of the object at TeV energies 
by MAGIC (M. Persic, priv. comm. and MAGIC collaboration, in prep.). 
Given that previous estimates place 1ES 0647+250 at z $>$ 0.3, 
if the detection is confirmed, this would make it one of the most distant 
TeV sources observed so far. Given the importance of 1ES 0647+250 to 
EBL studies, the evaluation of its distance is a fundamental measurement. 
Throughout this paper, we use the cosmology $H_0 = 70$ km s$^{-1}$ Mpc$^{-1}$, 
$\Omega_{M}$ = 0.3 and $\Omega_{\Lambda}$ = 0.7.

\section{Observations and data analysis}

1ES 0647+250 was observed at the 2.5m Nordic Optical Telescope (NOT), 
La Palma, during the night of November 12, 2008. We used the $1024\times1024$ 
pixel NOTCam detector in imaging mode with a pixel scale of 
0\farcs235 px${^-1}$, giving a field of view of $\sim 4\times4$ arcmin$^2$. 
The seeing during the observations, as derived from the median FWHM size of 
the stars in the frame, was $\sim$0\farcs6. The images in the $H$-band were 
acquired by dithering the target across the array in a random grid within 
a box of $\sim$20 arcsec, and taking a 60 sec exposure at each position, 
always keeping the target well inside the field. Individual exposures were 
then coadded to achieve the final frame. A total of 45 exposures of 60 sec 
were acquired, which provided a total exposure time of 2700 sec.  

\begin{figure}
\centering
\includegraphics[width=8.5cm]{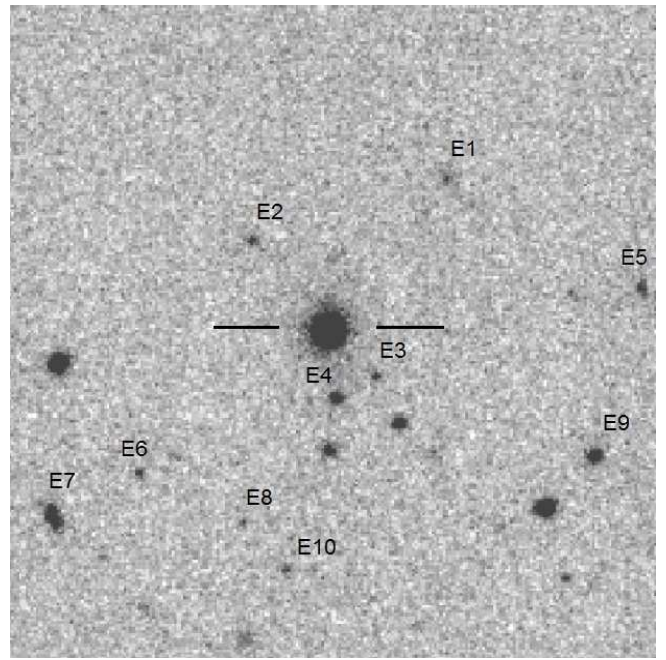}
\caption{\label{field} 
The $H$-band image of the close environment of 1ES 0647+250. 
The field size is 55 $\times$ 55 arcsec. 
North is up and east is to the left. 
1ES 0647+250 is the object between the bars in the middle, 
while the labelled objects 
are extended sources in the vicinity of 1ES 0647+250. 
}
\end{figure}

Data reduction was performed using IRAF\footnote{IRAF is distributed by the National Optical Astronomy Observatories, which are operated by the Association of Universities for Research in Astronomy, Inc., under cooperative agreement with the National Science Foundation.}. 
Bad pixels were corrected for in each image using a mask made from the ratio 
of two sky flats with different illumination levels. Sky subtraction was 
performed for each science image using a median averaged frame of all 
the other temporally close frames in a grid of eight exposures. Flat fielding 
was made using normalized median averaged twilight sky frames with different 
illumination levels. Finally, images were aligned to sub-pixel accuracy using 
field stars as reference points and combined after removing spurious 
pixel values to obtain the final reduced co-added image. Standard stars from 
Hunt et al. (1998) were observed temporally close to the target to provide 
photometric calibration, which resulted in internal photometric accuracy of 
$\sim$0.05 mag, as determined from the comparison of all observed 
standard stars. The final image of the field of 1ES 0647+250 is shown in 
Fig. \ref{field}.

Data analysis was carried out using AIDA 
(Astronomical Image Decomposition and Analysis), an IDL-based software package 
designed to perform 2D model fitting of quasar images, 
providing simultaneous decomposition into nuclear and host galaxy components. 
The applied procedure is described in detail in Kotilainen et al. (2007), 
and briefly summarized here.


The most critical part of the analysis is the determination of the PSF model 
and the estimate of the background level, which may strongly affect the faint 
signal from the diffuse object. 
To model the PSF shape, we used field stars selected 
on the basis of their FWHM, sharpness, roundness and S/N ratio, 
including bright, slightly saturated stars, in order to properly model 
the faint wing of the PSF. 

Each star was then modeled with four 2D Gaussians, representing the core of 
the PSF, and an exponential extended wing of the PSF. Regions contaminated by 
nearby sources, saturated pixels and other defects of the images were 
masked out. The local background was computed in a circular annulus centered 
on the star, and its uncertainty was estimated from the standard deviation of 
the values in sectors of concentric sub-annuli included in this area. 
Finally, the region used in the fit was selected by defining an internal and 
an external radius, allowing the exclusion of the core of bright, 
saturated stars. From the comparison of the resulting light profiles, we find 
no systematic variation of the PSF over the field of view. Thus, the same 
model was fitted simultaneously to all the stars of the image.

The uncertainty of the PSF model was estimated by comparing the analytical fit 
with the individual observed star profiles, and adding a fixed term 
($0.1$ mag arcsec$^-2$) to account for possible systematic effects in 
the data reduction (e.g. zero-point and PSF stability, and alignment in 
the de-jitter procedure). 


In order to evaluate whether the target is resolved, we first fit 
the image 
using only the PSF model, i.e. the unresolved nucleus. The residuals revealed 
a significant excess over the PSF shape beyond
$\sim$1 arcsec radius, where 
the contribution from the host galaxy becomes detectable, and the target was 
considered resolved. The target was then fitted with a two component model, 
consisting of the PSF and a host galaxy described by a Sersic profile, 
where the Sersic index was kept as a free parameter. 

\section{Results}

\begin{figure*}
\includegraphics[width=17cm]{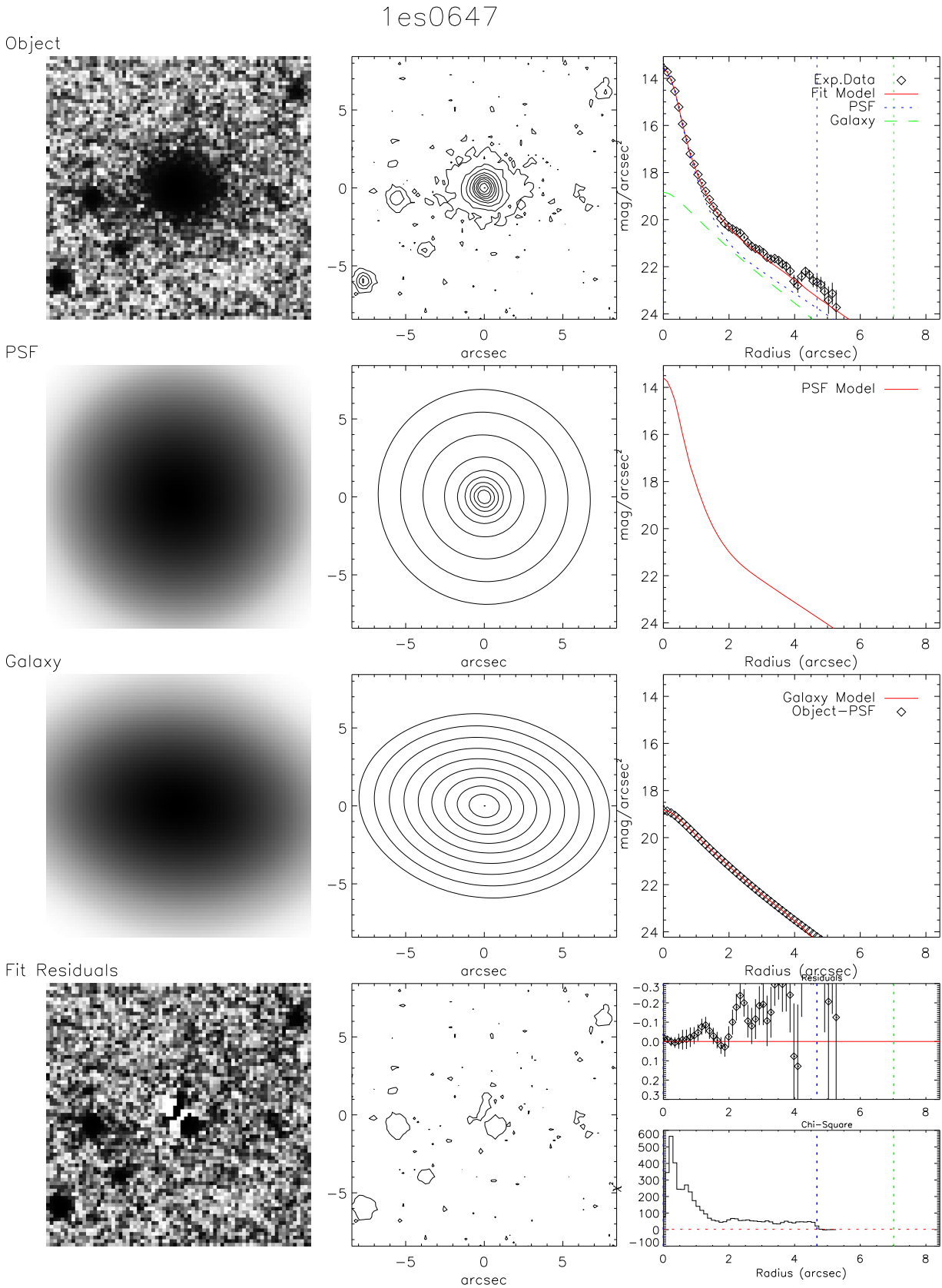}
\caption{\label{prof} 
The central 7 $\times$ 7 arcsec region surrounding 1ES 0647+250. In the 
left-hand and middle panels, in grey-scale and contours, respectively, 
from top to bottom: (a) the original image, (b) the PSF model, 
(c) the host galaxy (the PSF model subtracted from the observed profile), 
and (d) the residuals of the fit. 
On the right-hand panels, the top panel shows the observed radial profile 
(open diamonds), superimposed to the PSF model (blue dotted line) 
and an elliptical galaxy model convolved with its PSF (green dashed line). 
The red solid line shows the composite fit. The second and third panels show 
the radial profiles of the PSF model (red solid line), and the host galaxy 
after the PSF subtraction, respectively, while the bottom two panels show 
the residuals and the $\chi^2$ distribution of the fit. 
}
\end{figure*}

\begin{figure}
\includegraphics[width=8.5cm]{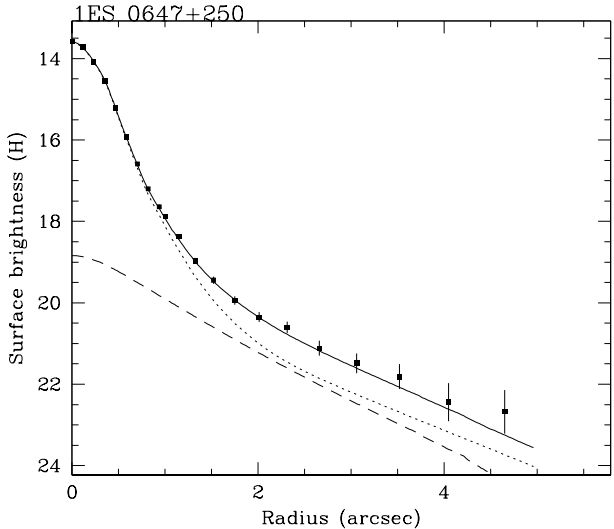}
\caption{\label{fit} 
The observed radial surface brightness profile (squares), superimposed to 
the PSF model (dotted line) and an elliptical galaxy model convolved with 
its PSF (dashed line). The solid line shows the composite fit. 
}
\end{figure}

In Fig. \ref{prof}, we report the image of 1ES 0647+250, the best-fitting PSF, 
the residual after the PSF subtraction (i.e. the host galaxy), 
and the residual after the fit with both the PSF and the galaxy model. 
Table \ref{results} gives the results of the model fit. 
The host galaxy of 1ES 0647+250 is faint ($H$ = 16.9) compared to the nucleus 
($H$ = 14.3) and never exceeds the core brightness at any radius, but is 
visible as a small excess in the surface brightness profile (Fig. \ref{fit}). 
The fit with the PSF + galaxy reduces the $\chi^2$ of the fit compared to 
the PSF only fit ($\chi^2$ = 1.9 and 2.6, respectively). 

\begin{table*}
\caption{\label{results} Model fit results.}
\centering
\begin{tabular}{lllllllll}
\hline
\hline
Model        & $\chi^2$ & m(nuc) & m(host) & $\mu_0$ & R(e) & ell   & $\Theta$ & N(Sersic)\\
             &          & (mag)  & (mag)   & mag arcsec$^{-2}$ & arcsec & & deg & \\
\hline
PSF only     & 2.6     & 14.21  & ...     & ...     & ...  & ...   & ...      & ...      \\
PSF + host galaxy       & 1.9     & 14.26  & 16.95   & 18.0    & 1.59 & 0.32 & 3.01     & 1.24     \\
\hline
\end{tabular}
\end{table*}

Knowing the host galaxy magnitude, we can estimate the redshift of 
1ES 0647+250. Sbarufatti et al. (2005) demonstrated that the distribution of 
the absolute magnitudes of BL Lac host galaxies is almost Gaussian with 
an average of $M_R$ = -22.8 and $\sigma$ = 0.5, and that BL Lac host galaxies 
can therefore be used as a standard candle to estimate their distances. 
Before using their method, we first have to transform the apparent 
$H$-band magnitude of the host galaxy to apparent $R$-band magnitude. 
Since the observed $R-H$ colour depends on redshift, we have to determine 
the redshift by iteration, starting from z = 0 and using Eq. (2) in 
Sbarufatti et al. (2005) and a typical BL Lac host galaxy colour 
$R-H$ = 2.2 $\pm$0.4 (Hyv\"onen et al. 2007). This iteration yields 
R(host) = 19.1 and z = 0.408 in the adopted cosmology.

To compute the uncertainty in the derived redshift, we use the estimated 
1$\sigma$ fitting uncertainty of the host galaxy magnitude of 0.2 mag. 
Performing the redshift iteration at m(host) + 0.2 mag and m(host) - 0.2 mag 
yields an error of $\pm$0.03 for z. By adding to this in quadrature 
the inherent uncertainty in the method ($\Delta z = 0.05$; 
Sbarufatti et al. 2005), we derive the final error in z to be $\pm$0.06. 

The effective radius of the host galaxy is 1\farcs6$\pm$0\farcs3, 
which translates into 8.6$\pm$1.7 kpc at z = 0.41, consistent with typical 
values found in the near-infrared (NIR) for blazar host galaxies 
(e.g. Kotilainen et al. 1998a,b), which lends further credibility to our 
estimate of the redshift. 

\section{Discussion}

There have been several previous attempts to derive the host galaxy properties 
of 1ES 0647+250 with varying success. Falomo \& Kotilainen (1999) found 
the object to be unresolved from 900 sec NOT $R$-band imaging taken under 
very good seeing conditions (0\farcs65). They derived a lower limit of 
z $>$ 0.3, assuming it is hosted by a typical elliptical galaxy. 
Scarpa et al. (2000) obtained a 600 sec HST F702W band image, as part of their 
BL Lac snapshot survey. They found the surface brightness profile to be 
consistent with the PSF and derived an upper limit R $>$ 19.1 for the apparent 
magnitude of the host galaxy. Based on this upper limit, 
Sbarufatti et al. (2005) derived a lower limit of z $>$ 0.47 for the redshift. 
Finally, Nilsson et al. (2003) found that 1ES 0647+250 was unresolved in 
a 900 sec NOT $R$-band image, taken under $\sim$1\farcs0 seeing conditions. 

Meisner \& Romani (2010) were able to resolve the host galaxy from 1200 sec 
$i'$-band imaging at the WIYN 3.6m telescope at Kitt Peak National Observatory, 
under very good seeing conditions (0\farcs68). They derived apparent $i'$-band 
magnitudes of 16.1 and 19.0$\pm$0.1 for the nucleus and the host galaxy, 
respectively. However, their data analysis did not allow them to 
estimate the effective radius of 
the host galaxy. Based on the host magnitude, and using the method of 
Sbarufatti et al. (2005), Meisner \& Romani (2010) report imaging redshift of 
z = 0.45$\pm$0.08.
We note that the detection of the host galaxy by Meisner \& Romani is not 
well documented, in the sense that they do not show the image of the field, 
the radial profiles vs. the models, nor the comparison with the PSF. 
Especially, the lack of an estimate of the effective radius translates into 
a poorer constraint on the redshift, than that reported in this work.

Knowledge of the redshift is of critical importance to the VHE spectra of 
blazars, due to the absorption of VHE photons via pair production on the EBL. 
This absorption is energy-dependent and increases strongly with redshift, 
distorting the VHE spectra of distant objects.
Our redshift estimate, z = 0.41$\pm$0.06, 
places thus
important constraints on the gamma-ray properties of 1ES 0647+250, 
and it is consistent with a picture where the VHE photons are 
mostly absorbed and the gamma-ray spectrum of 1ES 0647+250 is dominated by 
the EBL.


BL Lac hosts are often surrounded by a significant excess of nearby galaxies 
(e.g. Falomo et al. 2000). To see if there is an overdensity of galaxies 
around 1ES 0647+250, we counted all the resolved objects within a projected 
distance of 27\farcs5 (135 kpc at z = 0.41) of its nucleus, and in eight 
similarly sized comparison fields on the outskirts of the final image of 
the field. There are 10 resolved objects in the environment of 1ES 0647+250 
(see Fig. \ref{field}), whereas the corresponding average for 
the comparison fields is 6.4$\pm$3.0. This difference indicates that 
1ES 0647+250 may belong to a compact group of galaxies at z $\sim$0.4.
Interestingly, there are two close companions, namely galaxies E3 and E4 in 
Fig. \ref{field}. They are both at a projected distance of only 5\farcs5 
(27 kpc at z = 0.41), and have $H$-band magnitudes 20.3 and 18.6, 
respectively.

\section{Conclusions}

A deep NOT $H$-band image of the BL Lacertae object 1ES 0647+250 
has enabled us to detect and characterize the host galaxy in the near-infrared. 
Using the imaging redshift method by Sbarufatti et al. (2005), we estimate 
a redshift of z = 0.41$\pm$0.06 for 1ES 0647+250. This redshift is in 
agreement with 
constraints from its gamma-ray emission. 

We have shown that the imaging redshift technique by Sbarufatti et al. (2005) 
gives independently consistent results, and that imaging searches in 
the near-infrared even with a moderately sized telescope, such as the NOT, 
can be productive to considerably high redshifts. The contrast of 
the host galaxy with the blue nuclear continuum flux is at its maximum in 
the NIR, a fact that may allow host detections even in higher redshift sources 
with extreme core dominance. 
We encourage attempts with large 8-m class telescopes to secure 
the spectroscopic redshift of 1ES 0647+250 from optical spectroscopy. 

\begin{acknowledgements}

JKK and TH acknowledge financial support from the Academy of Finland, 
projects 8107775 and 2600021611. 
These data are based on observations made with the Nordic Optical Telescope, 
operated on the island of La Palma jointly by Denmark, Finland, Iceland, 
Norway, and Sweden, in the Spanish Observatorio del Roque de los Muchachos of 
the Instituto de Astrofisica de Canarias.  

\end{acknowledgements}

\end{document}